\documentclass[apjl]{emulateapj}

\shorttitle{Evolving Escape Fraction}
\shortauthors{Siana et al.}

\begin{document}

\title{A Deep HST Search for Escaping Lyman Continuum Flux at $z\sim1.3$:\\ Evidence for an Evolving Ionizing Emissivity$^{\dagger}$}

\author{\sc Brian Siana\altaffilmark{1}, Harry I. Teplitz\altaffilmark{2}, Henry C. Ferguson\altaffilmark{3}, Thomas M. Brown\altaffilmark{3}, Mauro Giavalisco\altaffilmark{4}, Mark Dickinson\altaffilmark{5}, Carrie R. Bridge\altaffilmark{2}, Ranga-Ram Chary\altaffilmark{2}, Duilia F. de Mello\altaffilmark{6}, Christopher J. Conselice\altaffilmark{7}, Jonathan P. Gardner\altaffilmark{8}, James W. Colbert\altaffilmark{2}, Claudia Scarlata\altaffilmark{2}}

\altaffiltext{1}{California Institute of Technology, MS 105-24, Pasadena, CA 91125}

\altaffiltext{2}{Spitzer Science Center, California Institute of Technology, 220-6, Pasadena, CA 91125}

\altaffiltext{3}{Space Telescope Science Institute, 3700 San Martin Drive, Baltimore, MD 21218}

\altaffiltext{4}{University of Massachusetts, Department of Astronomy, Amherst, MA, 01003}

\altaffiltext{5}{National Optical Astronomy Observatory, 950 N. Cherry Ave., Tucson, AZ 85719}

\altaffiltext{6}{Department of Physics, Catholic University of America, 620 Michigan Avenue, Washington DC 20064}

\altaffiltext{7}{University of Nottingham, Nottingham, NG7 2RD, UK}

\altaffiltext{8}{Astrophysics Science Division, Observational Cosmology Laboratory, Code 665, Goddard Space Flight Cent
er, Greenbelt, MD 20771}

\altaffiltext{${\dagger}$}{Based on observations made with the NASA/ESA Hubble Space Telescope, obtained at the Space T
elescope Science Institute, which is operated by the Association of Universities for Research in Astronomy, Inc., under
 NASA contract NAS 5-26555.  These observations are associated with program 10872.}

\begin{abstract}

We have obtained deep Hubble Space Telescope far-UV images of 15 starburst galaxies at $z\sim1.3$ in the GOODS fields to search for escaping Lyman continuum photons.  These are the deepest far-UV images ($m_{AB}=28.7$, $3\sigma$, 1$''$ diameter) over this large an area (4.83 arcmin$^2$) and provide the best escape fraction constraints for any galaxy at any redshift.  We do not detect any individual galaxies, with $3\sigma$ limits to the Lyman Continuum ($\sim 700$ \AA) flux 50--149 times fainter (in $f_{\nu}$) than the rest-frame UV (1500 \AA) continuum fluxes.  Correcting for the mean IGM attenuation (factor $\sim2$), as well as an intrinsic stellar Lyman Break (factor $\sim3$), these limits translate to relative escape fraction limits of $f_{esc,rel}<[0.03,0.21]$.  The stacked limit is $f_{esc,rel}(3\sigma)<0.02$.  We use a Monte Carlo simulation to properly account for the expected distribution of line--of--sight IGM opacities.  When including constraints from previous surveys at $z\sim1.3$ we find that, at the 95\% confidence level, no more than 8\% of star--forming galaxies at $z\sim1.3$ can have relative escape fractions greater than 0.50.  Alternatively, if the majority of galaxies have low, but non--zero, escaping Lyman Continuum, the escape fraction can not be more than 0.04.  In light of some evidence for strong Lyman Continuum emission from UV--faint regions of LBGs at $z\sim3$, we also stack sub--regions of our galaxies with different surface brightnesses, and detect no signficant Lyman continuum flux at the $f_{esc,rel}<0.03$ level.  Both the stacked limits, and the limits from the Monte Carlo simulation suggest that the average ionizing emissivity (relative to non-ionizing UV emissivity) at $z\sim1.3$ is significantly lower than has been observed in Lyman Break Galaxies (LBGs) at $z\sim3$.  If the ionizing emissivity of star--forming galaxies is in fact increasing with redshift, it would help to explain the high photoionization rates seen in the IGM at $z>4$ and reionization of the intergalactic medium at $z>6$.  Finally, we see no evidence of shadowing of the far--UV background by H{\sc i} in our target galaxies (on scales smaller than $\sim40$ kpc), suggesting that either the majority of the far--UV background originates at lower redshift (or within our galaxy) or that the high column-density H{\sc i} in our target galaxies  is distributed over larger than 40 kpc scales.  \\
 
\end{abstract}


\section{Introduction}

The H{\sc i}--ionizing background and its evolution are critically important to various aspects of galaxy assembly: reionization of the intergalactic medium (IGM) by $z\sim6$ \citep{fan02}; supression of star formation in low mass dark matter halos ($<10^8$ M$_{\odot}$) \citep{rees86,efstathiou92,bullock00,somerville02}; and affecting the H{\sc i} distribution in outskirts of galaxies \citep{dove94}.  Unfortunately, the sources of the ionizing background at high redshift are not yet known.  Searches for high redshift ($z>3$) QSOs show that their space densities fall precipitously toward higher redshift and are too rare at the highest redshifts to contribute significantly to the HI--ionizing background \citep{richards06,hopkins07,siana08,jiang08,jiang09}.  Therefore, young star-forming galaxies have become the most likely candidates for providing the ionizing emission.  However, it isn't clear what fraction of the ionizing photons (hereafter called the escape fraction, $f_{esc}$) produced by massive stars can escape the high column density H{\sc i} gas surrounding these star-forming regions.  Many studies have been conducted searching for escaping ionizing radiation from galaxies at various redshifts, with most providing only upper limits to the ionizing emissivity.  \citet{steidel01} reported the first detection of Lyman continuum (LyC) photons in a stack of 29 $z\sim3.4$ LBG spectra.  The flux ratio between the rest-frame 1500\AA\ and the Lyman continuum (corrected for average IGM absorption), $f_{1500}/f_{900}$, implies that the majority of Lyman Continuum photons produced by the massive stars are escaping into the IGM.  Deeper spectra of 14 LBGs show that most have low relative escape fractions ($f_{esc,rel}<0.25$), with a small fraction (2/14) exhibiting escape fractions near unity \citep{shapley06}.  In addition, \citet{iwata09} have conducted narrow-band imaging of the Lyman continuum of 125 Ly$\alpha$ emitters (LAEs) and 73 Lyman break galaxies (LBGs) at $z=3.09$ and detect $\sim10$\% with large escape fractions.  

While many galaxies at $z\sim3$ have now been detected in the Lyman continuum, all searches at low redshift ($z\sim0$ and $z\sim1$)  have given null results \citep{leitherer95, deharveng01, giallongo02, malkan03, siana07, grimes07, cowie09,grimes09}.  The most significant results have come from space-based far-UV imaging with HST and GALEX.  \citet{siana07} compiled \citep[along with data from][]{malkan03} a sample of 32 galaxies at $1.2<z<1.5$ with deep HST far-UV images with no significant detection of escaping Lyman continuum.  This implies that the galaxies must, on average, have a UV-to-LyC continuum ratio of  $f_{1500}/f_{LyC}(3\sigma)>50$.  Recently, \citep{cowie09} stacked GALEX far-UV images of 626 star-forming galaxies at $0.9<z<1.4$ and did not get a signficant detection, with a limit of $f_{1500}/f_{LyC}(3\sigma)>83$.  These studies suggest that either the average escape fraction has been decreasing with time, or that analogous galaxies at low redshift have not yet been targeted.   However, given the small fraction of galaxies at $z\sim3$ with observed large escape fractions ($\sim$1/10), we still would only expect a few detections thus far at $z\sim1$ if their escape fractions are similar to higher redshift galaxies.  Thus, larger samples are needed at low redshift.  

Most published escape fraction limits at $z\sim3$ are not very deep, with typical $f_{esc,rel}(3\sigma) \sim 0.25$.  Therefore, it is possible that many of the galaxies with non-detections do emit significant ionizing continuum below current detection limits.  Given that there are 10 times more objects with non-detections, even small amounts of ionizing emission from these galaxies would contribute significantly to (or even dominate) the ionizing background.  Therefore, deep surveys probing low escape fractions are also important in determing the total contribution of star formation to the ionizing background.  

In addition to simply measuring the escape fraction and its evolution, it is important to understand {\it how} these photons are able to escape into the IGM, given the presumably large reservoirs of gas and dust in these actively star-forming galaxies.  Several mechanisms have been invoked to explain high escape fractions, each of which predicts unique observational signatures.   Both semi-analytic models and numerical simulations suggest that SNe winds may be sufficiently powerful to expel gas from the disk, producing low H{\sc i} column density lines of sight (LOSs), though it is highly dependent on the spatial distribution of the star clusters \citep{clarke02} and the star formation efficiency \citep[relative to the gas mass,][]{fujita03}.  In such a scenario, escaping Lyman continuum may be expected in only the regions of the galaxy with high star formation surface density.  The simulations of \citet{razoumov06, razoumov07, razoumov09} suggest that feedback is far more efficient in lower mass halos.  However, \citet{gnedin08a} suggest that the smaller gas-to-stellar scale height ratios in high mass disks results in a higher probability of young stars being exposed to low column LOSs \citep[see also][for high redshift dwarf galaxy results]{wise09}.   In this scenario, we would expect to see escaping LyC photons from above (in azimuth) an edge on disk and only in the most massive galaxies.  Finally, \citet{ricotti02} posits that it is the formation of globular clusters (GCs) that may have reionized the universe.  In which case, we would expect to see no evidence for large escape fractions at $z\sim1$, as the vast majority of GCs had been formed long before then and should no longer contain massive stars.

Determining from which parts of the galaxies the photons are escaping can help explain the mechanisms by which such large fractions of Lyman continuum photons can escape these large H{\sc i} reservoirs.  Does the ionizing emission simply trace the UV (1500\AA) emission, or is it only escaping from certain portions of the galaxy?  What are the spectral energy distributions (SEDs) of the LyC emitting regions?  And if the LyC is only escaping from a portion of the galaxy, how can such large ionizing fluxes observed in some LBGs be emitted from only small fractions of the total starburst?

With these questions in mind we conducted a deep, high spatial resolution, far-UV imaging program of the 15 most UV-luminous star-forming galaxies at $1.2<z<1.5$ in the Great Observatories Origins Deep Survey \citep[GOODS,][]{giavalisco04}.  We use the Solar Blind Channel (SBC) of the Advanced Camera for Surveys (ACS) on the Hubble Space Telescope (HST) with the F150LP filter, which samples the rest-frame Lyman Continuum ($\sim700$ \AA) at these redshifts.  These data are typically a factor of two deeper than our previous survey \citep{siana07} and target galaxies about a magnitude brighter (at $\lambda_{rest}\sim1500$ \AA), giving a factor of 5--10 higher sensitivity to escaping LyC.  Also, the high spatial resolution allows us to investigate variations in the escape fraction via measurement of the relative distribution of the Lyman continuum and rest-frame UV fluxes to help determine the mechanism by which the Lyman continuum photons are allowed to escape.  

Throughout the text we use a flat $\Lambda$CDM cosmology with $H_{0}=70$ km s$^{-1}$ Mpc$^{-1}$, $\Omega_m=0.3$, and $\Omega_{\Lambda}=0.7$.  All flux densities are given in $f_{\nu}$ and all magnitudes are in the AB system.

\section{Observations and Data Reduction}

\subsection{Target Selection}

We targeted the 15 most UV-luminous galaxies in the GOODS fields with spectroscopic redshifts between $1.2<z<1.5$.  The sample has the brightest GOODS B-band (F435W, $\lambda_{rest}\sim1900$ \AA) fluxes as well as the bluest $B-I$ colors, in order to avoid dusty and/or older galaxies.  The specified redshift range keeps the Lyman limit redward of any significant system throughput ($z>1.2$) and ensures high sensitivity to the rest-frame UV continuum ($z<1.5$).  These galaxies have luminosities typical of the $L^*_{UV}$ LBGs at $z\sim3$ that exhibit large escape fractions.  The  targets were chosen in the GOODS fields for the wealth of ancillary data (HST, Spitzer, Chandra, optical spectroscopy, etc.) necessary for analyzing various galaxy properties (star formation histories, stellar masses, metallicities).  Finally, all X-ray sources were excluded to avoid contamination by unobscured AGN.

\subsection{Observations}

We imaged each galaxy with the Solar Blind Channel (SBC) of the Advanced Camera for Surveys (ACS) and the F150LP filter, a blue cutoff filter blocking light with $\lambda < 1450$ \AA.  This setup effectively blocks the geocoronal Ly$\alpha$ and [O{\sc i}] emission that would significantly increase the background in our images.  The red cutoff of the throughput is dictated by the decreasing sensitivity of the Multi-Anode Microchannel Array (MAMA) towards redder wavelengths.  The configuration gives very low throughput at $\lambda > 2000$ \AA\ ($<0.0008$ of peak sensitivity at 1500 \AA), with further decline in sensitivity at redder wavelengths.   We note here that the rate of decline in the QE of the MAMA at $\lambda>2000$ \AA\ has recently been found to be shallower than previously thought, resulting in a higher than expected QE at optical wavelengths.  However, we have performed tests with the new throughput curve for various SED scenarios and find that leakage of light redward of $\lambda>2000$ \AA\ is not a concern in this study as the expected leaking flux is below the noise limits achieved in this study.  The transmission weighted effective wavelength is $\lambda_{eff}=1610$ \AA, or $\lambda_{rest}\sim 700$\AA at $z\sim1.3$.  We note that because the photoionization cross-section falls as approximately $\nu^{-3}$, the HI opacity at $\lambda_{rest}\sim 700$ \AA\ is nearly half of that at 900 \AA.  

Each galaxy was imaged for five orbits, split into two visits of two and three orbits to ensure that SBC was not in use for too long, as the dark current increases as the instrument warms up  \citep[see][]{teplitz06}.  In addition, we also ensured that the visits did not follow any previous SBC use to ensure a low MAMA temperature.  Each orbit consisted of four dithered exposures of 630 seconds, giving a total exposure time of 12600 seconds per galaxy.  Finally, because the dark current glow is concentrated near the center of the detector, we placed the objects in the lower-left quadrant of the detector where the dark current glow is at a minimum.

\subsection{Data Reduction}

First we masked all known bad pixels and then subtracted the most recent dark image provided by Space Telescope Science Institute (STScI).  This dark image does not reflect the additional (and often much stronger) dark current ``glow'' that arises when the MAMA warms up.  We use our data to produce a profile for this dark ``glow'' to be subtracted from each image.  To do this we mask all known objects in the GOODS B-band images using the segmentation maps output from SExtractor \citep{bertin96} and add all of the exposures together.  A two dimensional 4th order spline is fit to the stacked dark, giving a smooth profile for the dark ``glow''. The amplitude of the ``glow'' changes with each exposure, so we scale the dark glow fit to the background of each image (with known objects masked) and subtract it.  The final, dark subtracted frames are then flattened using the flat field provided by STScI.  These reduced frames are drizzled \citep{fruchter02} onto the GOODS tiles to facilitate direct comparison with the other data ($pixfrac=0.6$, output pixel scale$=.03''$).  The relative astrometry between objects detected in both the SBC and GOODS B-band (F435W) images is less than one pixel (0.03$''$).  

In the region of the images where our primary targets are located, there is very little of the additional dark current glow, and the resulting drizzled images reach a $3\sigma$ sensitivity of $m_{AB}=28.7$ in a 1$''$ diameter aperture.  The regions near the center of the detector can be up to one magnitude less sensitive due to the additional dark current (which dominates the noise).  These data are the deepest images ever obtained at these wavelengths and over this large an area ($15\times0.32=4.83$ sq. arcmin).  

To define apertures for far-UV flux determinations, we started with the SExtractor \citep{bertin96} segmentation maps of the GOODS $z'$ (F850LP) images.  These apertures are much larger than the areas with significant emission at rest-frame UV wavelengths.  We therefore chose to shrink the apertures to isophotes that contain 90\% of the GOODS B-band (F435W).  This reduces the aperture sizes by about 50\%, resulting in sensitivities about 0.4 mags deeper.  Using the B-band data to determine the aperture is reasonable, as it is unlikely that a source emitting strongly at $\lambda_{rest} \sim 700$ \AA\ is not also emitting strongly at $\lambda_{rest} \sim 2000$ \AA.  

Finally, we have corrected the far-UV measurements for Galactic extinction using the estimates from the 100 $\mu$m COBE/DIRBE maps \citep[$A_V=$0.04 and 0.026 in GOODS-N and GOODS-S respectively,][]{schlegel98}.  The extinction curve of \citet{cardelli89} with an $R_V=3.1$ gives an $A_{1610}/A_{V} = 2.55$, resulting in $A_{1610} = $0.10 and 0.07 mags, respectively.  As these fields are some of the lowest extinction site lines out of the Galaxy, the extinction corrections are less than 10\%, even in the far-UV, and any uncertainties in the exact shape and normalization of the extinction curve do not significantly affect the photometry.  

\section{Results}

\subsection{Individual Limits}

\begin{figure*}
\epsscale{1.0}
\plotone{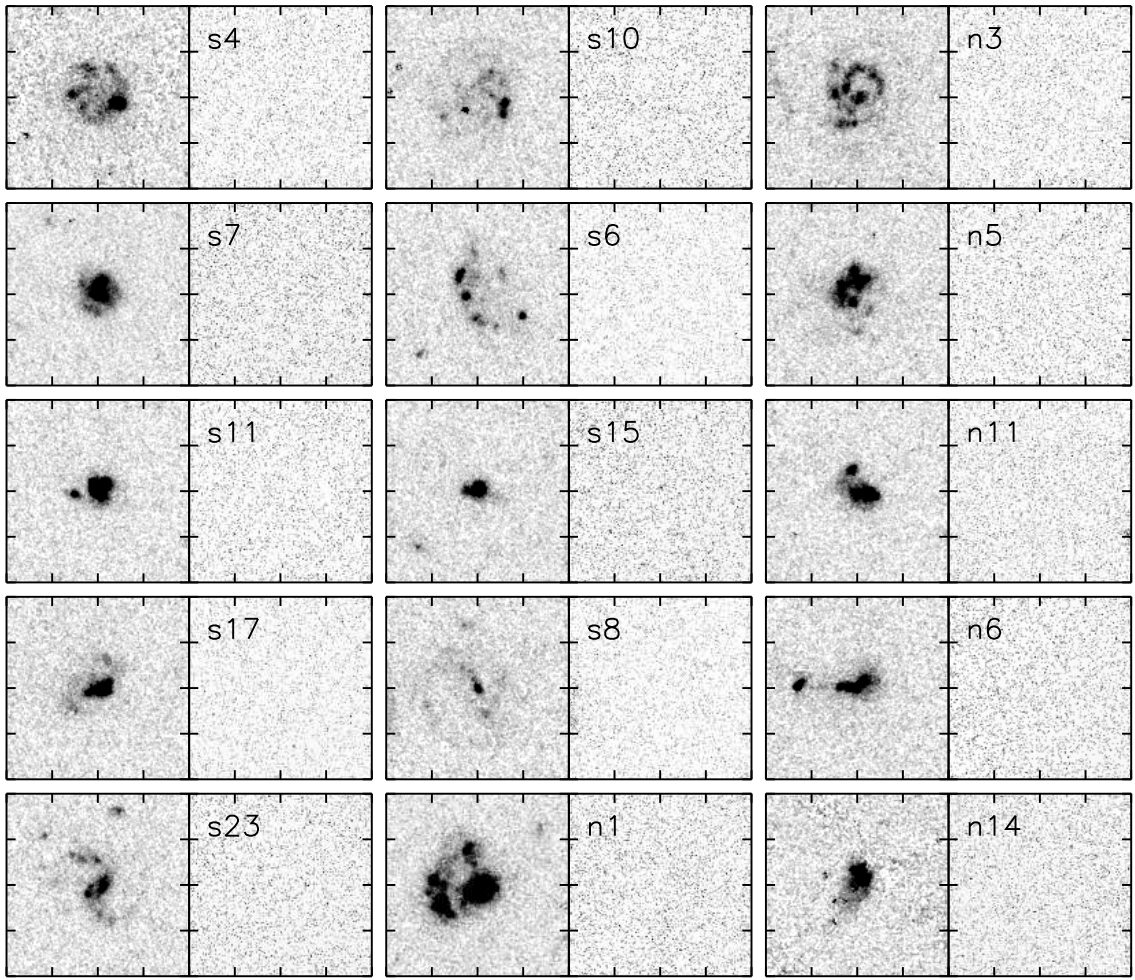}
\caption{GOODS B-band (F435W, left) and far-UV (F150LP, right) images of all 15 targets.  The stamps are 4$''$ on a side (1$'' = 8.4$ kpc at $z=1.3$).  The target galaxies display a variety of morphologies: including spirals, compact galaxies, mergers, and/or clumpy disks. \label{fig:stamps}}
\end{figure*}

\begin{figure}
\epsscale{1.0}
\plotone{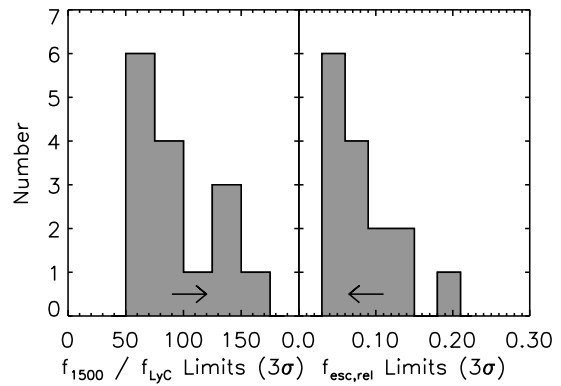}
\caption{Histogram of the $3\sigma$ limits to the UV-to-LyC ratio, $f_{1500}/f_{LyC}$, (left) and the resulting upper limits to the ionizing escape fraction (right).  To determine the relative escape fraction limits an intrinsic Lyman break of 3 (in $f_{\nu}$) was assumed and a redshift dependent IGM correction (factor of $\sim2$) was applied. The arrows indicate the direction of the limits, and emphasize that these are not detections. \label{fig:fesc_hist}}
\end{figure}

The final far--UV images of the galaxies are shown in Figure \ref{fig:stamps} alongside the GOODS B-band (F435W) images. None of the target galaxies were detected in the SBC far-UV images.  \citet{iwata09} see significant offsets between the detected LyC and the rest-frame UV continuum (typical offset $\sim 1 "$, or $\sim 7.6$ kpc at $z=3.1$).  Therefore, we have also searched for escaping Lyman continuum within a 1.5$''$ radius of our targets.  No detections are seen by eye.  In Table \ref{tab:info} we list the $3\sigma$ limits to the Lyman continuum flux, $f_{LyC}$, and the UV--to--LyC continuum flux ratio limits, $f_{1500}/f_{LyC}$. The $3\sigma$ limits to the UV--to--LyC flux ratios (in $f_{\nu}$) are shown in Figure \ref{fig:fesc_hist} and range from 50--174.  The depths of the images are nearly uniform, so the large range in UV--to--LyC limits is caused by the factor of 3-4 variations in both aperture area and 1500 \AA\ flux.  Before comparing these $f_{1500}/f_{LyC}$ limits to other surveys, it is important to correct for the average IGM opacity \citep[factor of $\sim2$,][]{siana07,inoue08} as this is a strong function of both redshift and wavelength.  These IGM-corrected limits, $(f_{LyC}/f_{1500})_{cor}$, can now be directly compared to other IGM-corrected detections and limits of other surveys.  We choose to use this ratio for comparison, rather than an estimate of the escape fraction, as there are additional assumptions made when determining the escape fraction.  However, for comparison with other surveys, we will also convert this to a ``relative'' escape fraction, which is defined as the fraction of LyC photons escaping the galaxy divided by the fraction of escaping 1500 \AA\ photons.  The relative escape fraction is computed as

\begin{equation}
f_{esc,rel} = \frac{(f_{1500}/f_{LC})_{stel}}{(f_{1500}/f_{LC})_{obs}}exp(\tau_{IGM})
\end{equation}

\noindent where $\tau_{IGM}\sim0.6$ is the LyC optical depth of the IGM at $z=1.3$ \citep[as measured through the F150LP filter,][]{siana07} and $(f_{1500}/f_{LC})_{stel}=3$ is the intrinsic Lyman Continuum break of the stellar population (in $f_{\nu}$).  In \citet{siana07} we argue that given typical metallicities and star formation histories for these galaxies, the intrinsic break may be much larger ($6-10$) but use this value to be consistent with the values assumed in other works.

\begin{center}
\begin{deluxetable*}{lrrccccccc}
\tabletypesize{\scriptsize}
\tablecaption{Target Information \label{tab:info}}
\tablehead{\colhead{Name} & \colhead{RA (J2000)} & \colhead{Dec (J2000)} & \colhead{$z_{spec}$} & \colhead{$U$} & \colhead{Area\tablenotemark{a}} & \colhead{$f_{1500}$\tablenotemark{bc}} & \colhead{$f_{LyC}$\tablenotemark{bd}} & \colhead{$f_{1500}/f_{LyC}$\tablenotemark{d}} & \colhead{$f_{esc,rel}$\tablenotemark{d}}}
\startdata
  s4  &  03:32:11.57 & -27:46:22.9 &   1.221 &   23.48 &  3.64 &  1.66 &  $<$0.0182 &   $>$90 &  $<$0.06 \\
  s7  &  03:32:13.66 & -27:43:13.1 &   1.856 &   23.46 &  1.50 &  1.64 &  $<$0.0138 &  $>$118 &  $<$0.21 \\
  s11 & 03:32:23.96 & -27:43:49.1 &   1.311 &   23.32 &  1.69 &  1.59 &  $<$0.0091 &  $>$174 &  $<$0.03 \\
  s17 & 03:32:37.37 & -27:50:13.7 &   1.389 &   23.26 &  2.06 &  1.85 &  $<$0.0144 &  $>$128 &  $<$0.05 \\
  s23 & 03:32:46.79 & -27:52:29.8 &   1.227 &   23.56 &  2.82 &  1.26 &  $<$0.0180 &   $>$69 &  $<$0.08 \\
  s10 & 03:32:22.92 & -27:48:57.4 &   1.264 &   23.47 &  4.02 &  1.41 &  $<$0.0208 &   $>$67 &  $<$0.08 \\
  s6  &  03:32:13.21 & -27:41:58.0 &   1.297 &   23.73 &  2.67 &  1.12 &  $<$0.0182 &   $>$61 &  $<$0.09 \\
  s15 & 03:32:35.25 & -27:54:32.6 &   1.414 &   23.66 &  1.22 &  1.22 &  $<$0.0082 &  $>$149 &  $<$0.05 \\
  s8  &  03:32:18.54 & -27:48:34.1 &   1.414 &   23.75 &  3.81 &  1.00 &  $<$0.0198 &   $>$50 &  $<$0.14 \\
  n1  & 12:36:12.42 &  62:14:38.4 &   1.432 &   22.90 &  5.17 &  2.96 &  $<$0.0200 &  $>$148 &  $<$0.05 \\
  n3  & 12:36:28.94 &  62:06:16.0 &   1.265 &   23.25 &  7.13 &  1.34 &  $<$0.0216 &   $>$61 &  $<$0.09 \\
  n5  & 12:35:55.87 &  62:13:32.7 &   1.296 &   23.25 &  3.12 &  1.52 &  $<$0.0164 &   $>$92 &  $<$0.06 \\
  n11 & 12:37:05.59 &  62:17:17.8 &   1.250 &   23.60 &  1.95 &  1.24 &  $<$0.0125 &   $>$98 &  $<$0.06 \\
  n6  & 12:36:45.71 &  62:07:54.3 &   1.433 &   23.92 &  4.09 &  1.04 &  $<$0.0207 &   $>$50 &  $<$0.14 \\
  n14 & 12:37:09.12 &  62:11:28.6 &   1.341 &   23.63 &  1.92 &  1.19 &  $<$0.0148 &   $>$80 &  $<$0.08 \\
\hline
Stacks \\
\hspace{0.1cm} High SB \tablenotemark{e} & \nodata &  \nodata &  \nodata  &  \nodata  &  \nodata  &    9.31 & 0.0199  & $>469$& $<$0.013 \\ 
\hspace{0.1cm} Low SB \tablenotemark{e} &  \nodata  &   \nodata &  \nodata  &  \nodata  &   \nodata  &  11.5 & 0.0575 & $>200$& $<$0.030 
\enddata
\tablenotetext{a}{arcmin$^2$}
\tablenotetext{b}{$\mu$Jy}
\tablenotetext{c}{0.9$\times$ the total value (to properly match far-UV aperture)}
\tablenotetext{d}{Limits are $3\sigma$}
\tablenotetext{e}{SB $=$ surface brightness, high (low) SB stacked above (below) $S_{B}(AB) = 23.06$ mag arcsec$^{-2}$}
\end{deluxetable*}
\end{center}

\subsection{Stacked Image}

\begin{figure}
\epsscale{1.0}
\plotone{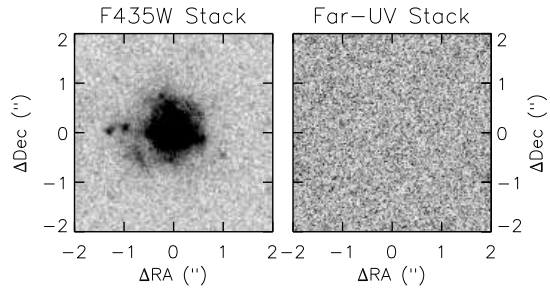}
\caption{The B-band (F435W, left) and far-UV (F150LP, right) stacked images.  The images are a simple sum of the individual objects.  No detection is seen in the far-UV image. \label{fig:stacks}}
\end{figure}

We have stacked both the F435W and FUV images of the 15 targets with a simple addition of all of the images (See Figure \ref{fig:stacks}).  No obvious detection is seen in the stack.  Of course, due to the varying sizes and morphologies of these galaxies, some galaxies are adding noise to areas where other galaxies have strong B-band flux.  Therefore, we do an optimized stacking which consists of adding up only the pixels that were used in the individual aperture extractions (based on the GOODS F435W isophotes).   We sum all of the pixels in the individual galaxy apertures (and their associated errors in quadrature) and get a total LyC flux of $f_{LyC}=2.2\times10^{-31} \pm 2.0\times10^{-31}$ erg s$^{-1}$ cm$^{-2}$ Hz$^{-1}$, not a significant detection. The limit to the UV-to-LyC ratio is $f_{1500}/f_{LyC} > 302$ ($3\sigma$).  After correcting for the average IGM transmission, this gives a $3\sigma$ limit to the LyC flux 151 times fainter than the UV flux.   

\begin{figure}
\epsscale{1.0}
\plotone{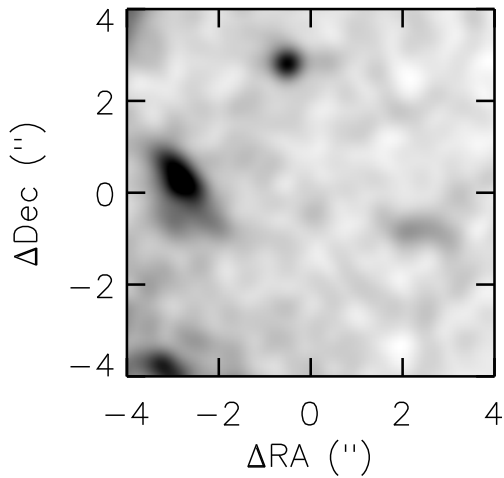}
\caption{Same far-UV (F150LP) stacked image as in Figure \ref{fig:stacks}, but zoomed out by a factor of two and smoothed with a 0.5$''$ Gaussian.  No clear detection exists at the center, nor is there a ``shadow'' of the background by our target galaxies.  Detected galaxies have not been subtracted in order to give an idea of the low level of confusion. \label{fig:stack_smooth}}
\end{figure}

\citet{cowie09}, in their GALEX far-UV stack of 626 galaxies, got a $\sim2\sigma$ negative flux in the location of the stack relative to the background.  Though it was unclear whether this may have been a systematic effect, it was suggested that a negative flux could be caused by a shadowing of the far-UV background by the H{\sc i} in the target galaxies.  In our stacked image we see no evidence for a flux below the background level.  In Figure \ref{fig:stack_smooth} we show a zoomed out version of our far-UV stack smoothed with a 0.5$''$ Gaussian.  There is no evidence for a shadow in this image (with scales less than $5"$, or 40 kpc).  The shadow seen by \citet{cowie09} appears to be large (5-10$''$), but the signal is very low, and this is of order the size of the GALEX far-UV beam so it is unclear whether the absorption is real or over what angular scale the absorption is taking place.  The lack of shadow seen in our data is perhaps not surprising, as most of the UV background \citep{brown00} likely originates in the foreground of these galaxies, either from Galactic sources \citep{hurwitz91} or Ly$\alpha$ emission from foreground galaxies \citep{henry91}. 

It is still not known which parts of galaxies are likely to have high LyC escape fractions, or what physical properties dictate the escape fraction.  One property that may affect the escape fraction is the star-formation surface density.  On one hand, regions of high star-formation surface density may provide sufficient winds to expel gas and produce low H{\sc i} column density lines-of-sight.  Alternatively, regions of low star formation surface density (like the far outskirts of disks in the local universe) have lower H{\sc i} columns so feedback may therefore be more effective at disrupting the gas distribution.  Indeed, \citet{iwata09} suggest that many of their LyC detections are offset by several kiloparsecs from the brightest regions of the galaxies.  It is unclear whether this emission is coming directly from star-formation or from nebular free-bound continuum emission from ionized gas on the outskirts of the galaxy \citep{inoue09} .  To determine if such extended LyC emission exists in our galaxies, we first checked all of the individual images by eye to make sure that our apertures (defined by the rest-frame 1900 \AA\ continuum) were not missing any LyC flux with extreme offsets ($>1''$).  Second, within our previously defined B-band apertures, we chose to separate regions with low and high UV surface brightness (dividing the sample at $S_{B}(AB) = 23.06$ mags arcsec$^{-2}$, the typical 10$\sigma$ sensitivity of a single pixel in the GOODS B-band images (corresponding to a star formation surface density of 0.187 $M_{\odot}$ yr$^{-1}$ kpc$^{-2}$).  Both the high and low surface brightness regions give non-detections in the LyC.  $(f_{1500}/f_{LyC})_{high-sb}> 469$ ($3\sigma$) and $(f_{1500}/f_{LyC})_{low-sb}>200$ ($3\sigma)$.  Note that we do not have resolved U-band images of these galaxies so, for these surface brightness stacks, we used the B-band fluxes and multiplied by 0.85 (the average flux ratio between the U and B images) to estimate $f_{1500}$.  The details of these stacked limits are also provided in Table \ref{tab:info}.  

\section{Comparison with Studies at $z\sim3$}

\begin{figure*}
\epsscale{1.0}
\plotone{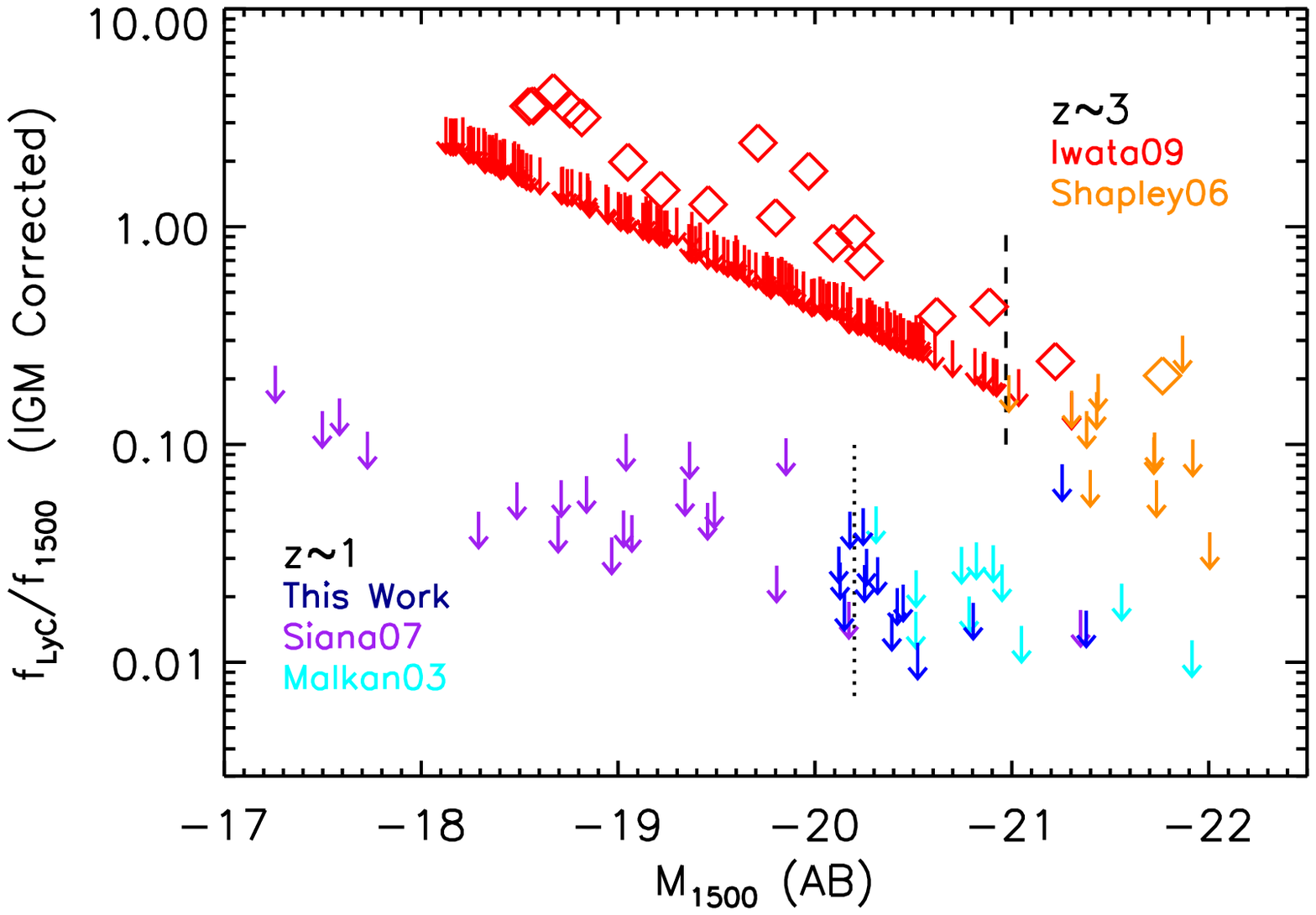}
\caption{Our new $f_{LyC}/f_{1500}$ measurements with other individual limits at $z\sim1$ \citep{malkan03, siana07} and $z\sim3$ \citep{shapley06,iwata09}.  The vertical dashed and dotted lines represent $M^*_{1500}$ at $z\sim1$ \citep[-20.2,][]{arnouts05}, and $z\sim3$ \citep[-20.97,][]{reddy08}, respectively.  The studies at both redshifts sample similar absolute magnitudes.  All reported limits have been adjusted to $3\sigma$ and each limit has been corrected to account for the mean IGM transmission at each redshift.  Nearly all of the $z\sim1$ measurements have $f_{LyC}/f_{1500}$ limits below the observed ratios of $z\sim3$ galaxies with LyC detections.}

\label{fig:fesc_meas}
\end{figure*}

In Figure \ref{fig:fesc_meas} we show the $f_{LyC}/f_{1500}$ measurements (versus UV luminosity) from several large surveys at $z\sim1$ and $z\sim3$ corrected for the average IGM attenuation at the relevant wavelengths ($\lambda_{rest}\sim700$ \AA\ at $z\sim1.3$ and $\lambda_{rest}\sim900$ \AA\ at $z\sim3$).

At $z\sim3$, recent studies have shown large LyC-to-UV ratios ($f_{LyC}/f_{1500}\geq0.2$) in $\sim$10\% of galaxies \citep{shapley06,iwata09}.  However, there are no LyC detections at $z\sim1$ despite the large number of galaxies (47) that have $3\sigma$ upper limits to $f_{LyC}/f_{1500}$ that are lower than any ratios seen in $z>3$ galaxies with LyC detections.  

It appears evident that a higher fraction of galaxies have large escape fractions at $z\sim3$ than at $z\sim1.3$.  However, it is first important to determine if these galaxies have similar stellar populations, as differences in the stellar populations will affect the intrinsic LyC fluxes (and the implied escape fractions).  Also, the characteristics of the foreground IGM are different in the two samples for two reasons:  the number of absorbers (per H{\sc i} column density) drops dramatically between $z\sim3$ and $z\sim1$ and different rest-frame wavelengths are probed at the two redshifts (700 \AA\ at $z\sim1.3$ and 900 \AA\ at $z\sim3$).  Therefore, in order to properly compare the samples at $z\sim3$ and $z\sim1$, we first need to determine that the two galaxy populations have similar stellar populations (and thus intrinsic $f_{LyC}/f_{1500}$ ratios), and then determine the effects of the IGM through Monte Carlo simulations based on the known redshift and column density distributions of H{\sc i} absorbers in the IGM.

\subsection{Galaxy Characterization}

To derive an escape fraction from the $f_{LyC}/f_{1500}$ limits, we must first have an estimate of the intrinsic $f_{LyC}/f_{1500}$ ratios.  Specifically, we must be certain that the stellar populations of these galaxies have strong intrinsic Lyman Continuum emissionand have similar properties to the $z\sim3$ LBGs that have been shown to exhibit high escape fractions.  To these ends, we fit the \citet{bruzual07} stellar population models to the available optical (ground-based $U$, GOODS HST B,V,I,z), near-IR, and $Spitzer$ 3.6 and 4.5 $\mu$m photometry to estimate the intrinsic LyC flux, stellar mass, starburst age, and dust reddening for each galaxy.  

For each galaxy we assume a Salpter IMF \citep{salpeter55}, a Calzetti reddening curve \citep{calzetti97}, and solar metallicity.  We fit to three different exponentially declining star formation histories ($SFR \propto e^{-t/\tau}$) with $e$--folding times of $\tau=$0.1 and 0.3, and $\infty$ (constant SFR)  Gyr.  The ages, star formation rates, and dust reddening are allowed to vary.  

\begin{figure}
\epsscale{1.0}
\plotone{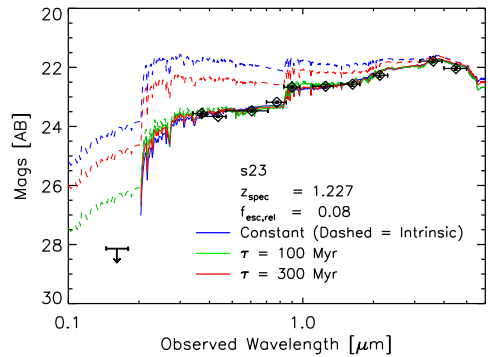}
\caption{Stellar population model fits to the optical, near-IR, and mid-IR photometry.  Three star formation histories are used (constant star formation and exponentially declining star formation with $e$-folding times of 100 \& 300 Myr.  The amplitude of the Balmer break is very small, indicating young $<300$ Myr old starbursts.  Note that, although the fits with different SEDs give very similar $\chi ^2$ values, the predicted intrinsic Lyman Continuum varies significanty (typically by a factor of $\sim 5$ in our fits).}
\label{fig:sedfits}
\end{figure}

Though the $S/N$ of the photometry is high and spans the rest-frame UV, optical and near-IR, the precise star-formation history is still ambiguous.  That is, the $\chi^2$ values are similar for the best-fit SEDs to the three different star-formation histories.  This is not too surprising, as the effect of the subtle differences in star-formation history can be masked by varying reddening and age.  Unfortunately, the best-fit SEDs using the three different SFHs give very different predictions of the intrinsic Lyman Continuum flux.  This is because the portion of the spectrum to which we are fitting our photometry is dominated by stars with ages $>10$ Myr (typically greater than 100 Myr), whereas the Lyman Continuum flux is dominated by massive stars that have lifetimes shorter than 10 Myrs.  This is demonstrated in Figure \ref{fig:sedfits}, where the fits with three different assumed SFHs are plotted.  At wavelengths where we have photometry, the difference between the fits is negligible.  However the intrinsic SEDs (ie. not attenuated by dust or H{\sc i} in the IGM, shown as dotted lines) exhibit very different Lyman Continuum fluxes (with typical variatons of a factor of $\sim5$ for our galaxies).  However, we think it is important to note that a constant star formation history gives a maximum Lyman Continuum flux of any of the typically considered histories (single burst, constant, or exponentially decline) and using this SFH still yields Lyman-break amplitudes of $f_{1500}/f_{LC} \sim 8$, similar to what was found in \citet{siana07}.  In \citet{siana07} we fit to both the \citet{bruzual03} and Starburst99 models \citep{leitherer99} and got very similar Lyman-break amplitudes and stellar population parameters.

\begin{figure*}
\epsscale{1.0}
\plotone{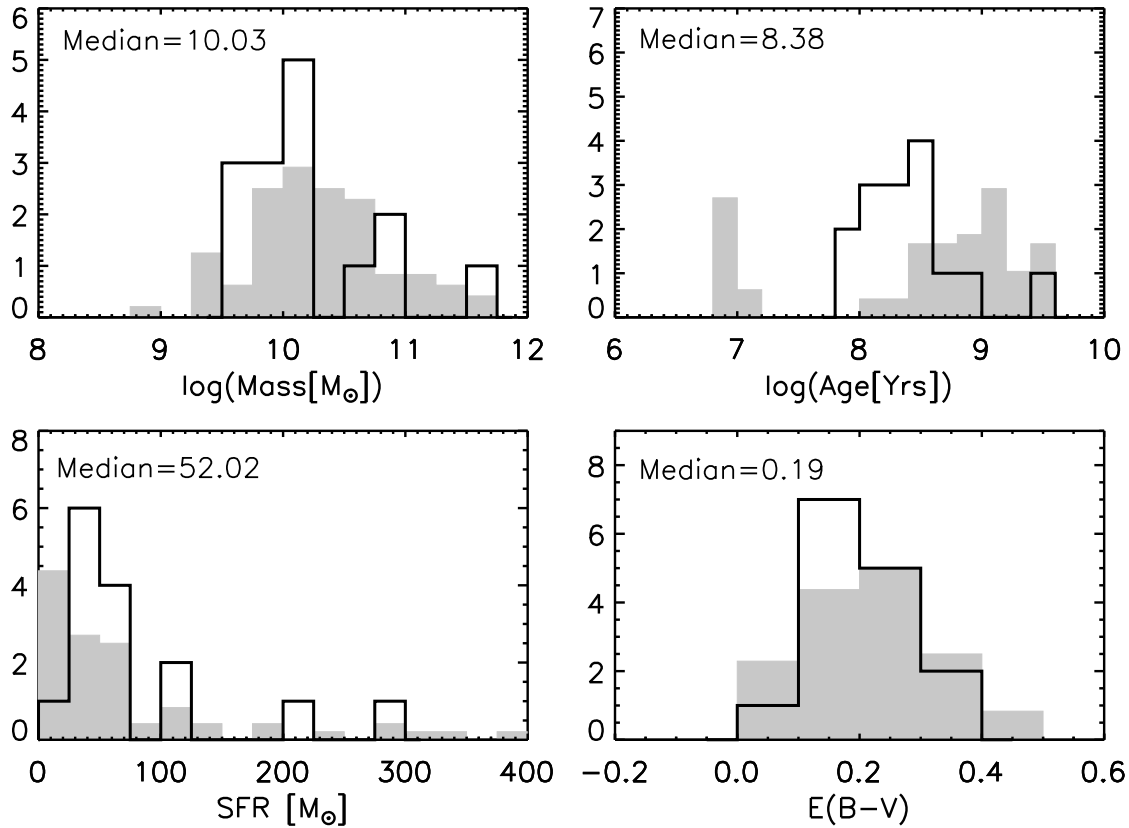}
\caption{Comparison of galaxy parameters for our $z\sim1.3$ sample (thick line) compared with the sample of \citep[][shaded histogram]{shapley05}.  For ease of comparison, the \citet{shapley05} histograms have been scaled so that they have the same normalization as our sample (actual sample contaions 72 galaxies).  Though our sample is small, it can bee seen that our galaxy sample has very similar properties to the higher redshift ($z\sim2$) LBG samples.}
\label{fig:comp_age_mass}
\end{figure*}

Regardless of how well we can predict the intrinsic LyC luminosities of these galaxies, the SED fitting is useful for comparing our sample to higher redshift galaxies selected via the Lyman Break technique.  In Figure \ref{fig:comp_age_mass}, we compare our best-fit parameters to those of a sample of 72 LBGs of \citet{shapley05}.  In both cases, a constant star formation history, Calzetti reddening law, Salpter IMF, and solar metallicity were assumed.  The distributions of dust reddening, SFR, and stellar mass are very similar to the distributions for the $z\sim2$ LBGs, though the best-fit stellar ages span a narrower range toward the low end of the LBG distribution (at $\sim 300$ Myr).   Given the similarities in the derived parameters, we are confident that any differences in the LyC-to-UV flux ratios between these two samples are not due to fundamental differences in the stellar populations.  

\subsection{Monte Carlo IGM Modeling}
\label{mc}

In Figure \ref{fig:fesc_meas}, we corrected for the average IGM attenuation of the LyC at the relevant wavelengths.  At $z\sim1.3$, the mean IGM transmission is about 0.5 but the distribution along different lines-of-sight is large \citep{siana07,inoue08}.  In 10--15\% of LOSs, the IGM is completely opaque and we would not expect to see any flux, regardless of the escape fraction.  Alternatively, 25-30\% of LOSs have transmission greater than 0.8 and our limits are far more sensitive than assumed in Figure \ref{fig:fesc_meas}.   In order to properly consider the large range in IGM optical depths towards galaxies at $z\sim1.3$, we have performed a Monte Carlo simulation to determine the escape fraction parameter space in which our galaxies must lie in order to get non-detections for every object.  The method is exactly the same as the method outlined in \citet{siana07}, but is briefly summarized here.  

In our simulations, we assume that all of these blue, UV-luminous galaxies have an intrinsic flux ratio (in $f_{\nu}$) of $f_{1500}/f_{LyC}=3$.  We assume a simplified parameter space for the relative escape fraction such that X fraction of galaxies have a relative escape fraction, Y, and the rest have $f_{esc,rel}=0$.  We step in X,Y parameter space in 50 evenly spaced intervals (between 0 and 1).  For each location in this parameter space, we randomly select 10,000 galaxies from this assumed distribution and place them along lines-of-sight with opacities randomly selected from the IGM distributions determined in \citet{siana07}.  We then determine what fraction of the time our experiment would produce only non-detections (as we observe).  The parameter space where we would obtain non-detections 84.1, 97.7, and 99.9\% of the time are considered our 1, 2, and 3$\sigma$ limits on the escape fraction parameter space.  We use the list of all the observed $f_{1500}/f_{LyC}$ $3\sigma$ limits from our observations, as well as those of \citet{malkan03} and \citet{siana07}.  

\begin{figure}
\epsscale{1.0}
\plotone{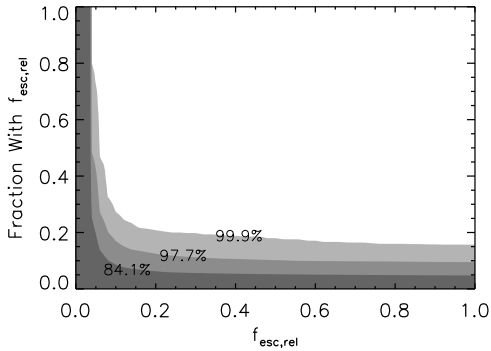}
\caption{Escape Fraction parameter space at $z\sim1.3$ allowed by our observations (in combination with limits from \citet{malkan03,siana07}).  In our simplified parameter space, a fraction of galaxies (y-axis) have a non-zero relative escape fraction (x-axis).  The darker shaded areas show the parameter space excluded at the 1, 2, and 3$\sigma$ levels from our Monte Carlo simulations.  At 95\% confidence level, we can rule out 8\% of galaxies having $f_{esc,rel}>0.5$ or, alternatively, rule out 50\% of galaxies having $f_{esc,rel}<0.04$.}
\label{fig:mc}
\end{figure}

The results of this Monte Carlo simulation are summarized in Figure \ref{fig:mc}.  As can be seen, the vast majority of the parameter space is ruled out with our observations.  At the 95\% confidence level there can be no more than 8\% of galaxies with $f_{esc,rel}>0.5$ or no more than 50\% of galaxies with $f_{esc,rel}>0.04$.  Both the increased depth of the survey and the larger sample help significantly increase the constraints over those achieved in \citet{siana07}.

Unfortunately, a similar analysis is difficult for LBGs and Ly$\alpha$--emitters at $z\sim3$.  First, there is a concern that some of the detections by \citet{iwata09} may actually be due to contamination from foreground sources.  Some of these LyC detections have large offsets (some greater than 10 kpc) from the UV--luminous LBGs with spectroscopic redshifts, raising concerns that the presumed LyC emission is actually non-ionizing emission from faint foreground galaxies.  

In addition, many of the LyC detected Ly$\alpha$ emitters (LAEs) in \citet{iwata09} display LyC fluxes (in $f_{\nu}$) equal to or greater than the rest-frame UV (1500 \AA) continuum, whereas stellar population models display a large decrement shortward of the Lyman limit.  \citet{inoue09} argue that this LyC ``bump'' may be due to free-bound continuum emission from the surface of dense H{\sc i} regions that are illuminated by star-forming regions with large escape fractions.  This mechanism would result in more energetic LyC photons being reprocessed by the H{\sc i} regions and re--emitted (via free-bound emission) at wavelengths just short of the Lyman Break.  If the electron temperature of the gas is fairly low ($\sim 10^4$ K), the thermal energy of the electron is insufficient to add significant energy to the resulting bound-free photons.  Thus, this LyC ``bump'' would not extend to wavelengths much below the Lyman limit ($\sim750$\AA).  This would naturally explain why we do not see such strong LyC flux with our observations, because we are most sensitive to LyC photons at lower rest-frame wavelengths (650-750 \AA).  However, we do not believe that free--bound nebular emission is significant in our sample at $z\sim1.3$, because rather unique conditions are required (extremely young ages ($<1$ Myr), low metallicities ($<$1/50 $Z_{\odot}$), and/or top--heavy IMFs), none of which should be prevalent in massive starbursts at $z\sim1.3$.  If some of the $z\sim3$ LBG  LyC detections are actually contaminants or if the 900 \AA\ LyC flux in LAEs is boosted via free-bound emission, this makes direct comparison with our results difficult, as it is unclear what the intrinsic LyC flux is in either case.

Generally, it appears that $z\sim3$ studies are finding 8--10\% of galaxies with large escape fractions \citet[1/14 in][]{shapley06} and \citet[7/73 LBGs and 10/125 LAEs in][]{iwata09}.   This is also the case in a much larger spectroscopic survey of $z\sim3$ LBGs, where $\sim$10\% are found with large escape fractions (M. Bogosavljevic 2009, private communication).  The results plotted in Figure \ref{fig:mc} show that, if 8\% of $z\sim1.3$ galaxies have large relative escape fractions ($>0.5$), then we should have seen at least one detection in the $z\sim1.3$ studies in 95\% of our IGM realizations.  

Finally, we note that in \citet{siana07} we do have a far-UV detection that is offset $\sim0.3''$ from the primary target galaxy and is associated with a faint object in the HST optical images.  If the object is not a foreground contaminant and is at the the same redshift as the target galaxy, $z=1.355$, the far-UV to optical SED would imply no flux decrement shortward of the Lyman limit and suggests significant ionizing emission that is significantly displaced from the primary galaxy, similar to some of the candidate Lyman continuum emitters found by \citet{iwata09}.  However, we have assumed that the souce is simply a faint foreground object. This example serves as a caution to LyC studies at higher redshifts, where the lines--of--sight are much longer and far more subject to contamination (Siana et al. 2007, Vanzella et al. 2010, in prep). 

\section{Discussion}

We have demonstrated that the average ionizing emissivity of star-forming galaxies is lower, relative to the non-ionizing UV emissivity, at $z\sim1$ than at $z\sim3$.  Our SED fits show that the stellar population parameters of our sample do not differ significantly from those of LBGs at $2<z<3$.  Therefore, there is no direct evidence that the increased LyC emission is due to an {\it intrinsic} difference in the LyC production from young stars.  If the stellar populations, and thus LyC productions, are the same, that would imply that the increased ionizing emission at high redshift is caused by an evolution in the escape fraction of ionizing photons due to differences in H{\sc i} masses and spatial distributions.  Of course, it is possible that the properties of a subset of the high redshift sample may have unique characteristics (top-heavy initial mass function, low metallicity) which would increase the intrinsic LyC-to-UV ratio, and appear to have a high escape fraction.  Regardless of whether the escape fraction is evolving or not, it appears the ionizing emissivity is increasing toward higher redshifts.  This has important implications for models of reionization and the evolution of the IGM.

An evolving LyC escape fraction has been inferred before.  \citet{inoue06} argue that at $z<2$, QSOs can provide all of the ionizing background (though the background estimates at low redshift are uncertain), so the LyC escape fraction from star-forming galaxies at $z<2$ is negligible.  However, because of the very low QSO space densities at $z>3$, they argue that star-forming galaxies at very high redshifts must provide the vast majority of the ionizing background, which implies higher escape fractions at higher redshifts.  Of course, this is an indirect argument, so it is important to confirm this assumption with a direct empirical detection.  

Recent investigations of the ionizing background at high redshift \citep{bolton05, becker07} note that the H{\sc i}--photoionization rate is fairly flat between $2<z<4$, which implies an increase in the ionizing emissitivity at higher redshifts \citep{faucher-giguere08}.  However, the total star formation rate density is actually declining towards higher redshifts \citet{bouwens07}.  This seeming contradiction can be resolved if in fact the LyC escape fraction is increasing towards higher redshifts.  

Some authors have pointed out the difficulty of ionizing the IGM by $z\sim11$ \citep[as is suggested by the WMAP5 electron scattering optical depth,][]{komatsu09} and maintaining that ionization given the relatively small average escape fractions, $<f_{esc,rel}> \sim 0.1$, exhibited at $z\sim3$ \citep{chary08,gnedin08b}.  If the escape fraction continues to increase with redshift beyond $z=3$, it would naturally remedy this discrepancy, eliminating the need to invoke top--heavy IMFs \citep{chary08} or more exotic mechanisms. 

It is interesting to compare the evolution in the escape fraction with predictions from numerical simulations.  This is a difficult problem for simulations, as the escape fraction is likely highly sensitive to the star formation and feedback presciptions.  \citet{gnedin08a} argue that the escape fraction is highest in high mass galaxies.  They see no evolution in the escape fraction, but they only follow their galaxies from $z=10$ to $z=4$, so it is not clear that they will see an evolving escape fraction at lower redshift.  \citet{razoumov06} get very different results, in that low mass galaxies have the highest escape fractions.  Furthermore, they see an evolving escape fraction over all redshifts covered \citep[$10<z<2.4$,][]{razoumov09}.  

Of course, it would be extremely useful to mesure the escape fraction at $z>6$, near the epoch of reionization.  However, it is impossible to directly determine if the ionizing emissivity of galaxies continues to increase at $z>3$, as the increasing IGM opacity makes direct LyC detection nearly impossible.  In the future, it will be important to identify unique characteristics of the LyC emitters at $z\sim3$ (at $\lambda_{rest}>1216$ \AA) and determine if galaxies with these characteristics are more common during the epoch of reionization ($z>6$).

\section{Conclusions}

We have performed a deep (5 orbits/galaxy) HST ACS/SBC far-UV imaging survey of 15 galaxies at $z\sim1.3$ to probe their rest-frame Lyman Continuum (700 \AA) emission.  The data achieve a depth of $m_{AB}(3\sigma)>28.7$ (AB) in a 1$''$ diameter aperture and are the deepest extragalactic far-UV images of comparable area.  Our findings are as follows:

\begin{itemize}

\item We do not detect any escaping Lyman Continuum from our target galaxies and achieve $3\sigma$ limits to the UV-to-LyC ratio of $f_{1500}/f_{LyC}>[50,149]$.   Our stacked image gives a $3\sigma$ limit of $f_{1500}/f_{LyC}>302$.  After correcting for average IGM opacity (factor of 2), and accounting for the intrinsic Lyman Break in the SED (factor of 3), these translate to individual relative escape fraction limits ($3\sigma$) of $f_{esc,rel}<[0.03,0.21]$ and $f_{esc,rel}<0.02$ in the stack.  These are the best escape fraction limits ever obtained at any redshift. 

\item Fits of stellar population models to the rest-frame UV to near-IR SEDs of our sample show them to have star formation rates, stellar masses, dust extinction, and ages similar to LBGs at higher redshift ($2<z<3$), some of which exhibit very high escape fractions.  We show that the intrinsic Lyman continuum flux is extremely difficult to determine as it depends on the star formation history within the last $<10$ Myr.  Therefore, it is impossible to infer the absolute escape fraction via SED fits alone and other means, such as H$\alpha$ flux measurements, are required.  Regardless of the exact value of the intrinsic Lyman continuum emissivity, the similarity in the stellar populations between our sample and LBGs suggests that the lack of LyC detections at $z\sim3$ is not likely due to a relative lack of Lyman continuum production, but rather a difference in the H{\sc i} mass and its distribution in and around galaxies at the two redshifts.

\item In order to properly account for varying opacity of the IGM along different lines-of-sight, we perform a Monte Carlo simulation to constrain the escape fraction with the limits from our data, as well as two similar surveys at $z\sim1.3$ \citep{malkan03,siana07}.  At the 95\% confidence level, there can be no more than 8\% of galaxies with very high escape fraction ($f_{esc,rel}>0.5$).  Alternatively, if most galaxies have a low, but non--zero, escape fraction, it must be less than 0.04.

\item The limits on the LyC-to-UV flux ratios of the 47 galaxies at $z\sim1.3$ from \citet{malkan03,siana07} and this work, are lower than  {\it any} of the LyC detections at $z\sim3$.  Generally, it appears that 8-10\% of $z\sim3$ galaxies exhibit large escape fractions ($f_{esc,rel} > 0.5$).  Our Monte Carlo simulations (which account for variations in IGM line-of-sight opacity) show that we would have observed at least one LyC detection 95\% of the time we perform our experiment.  Our lack of any detections in such a large sample strongly suggests that star-forming galaxies at $z\sim1$ have a lower ionizing emissivity than galaxies of comparable luminosity at $z\sim3$.

\item In light of recent claimed detections of escaping LyC from areas of low UV surface brightness, we have stacked subregions of the galaxies based on surface brightness limits (above and below an observed B-band surface brightness of $S_{AB}=23.04$ mag arcsec$^{-2}$.  We do not detect escaping LyC in either sample, with limits of $f_{esc,rel}(high)<0.013$ and $f_{esc,rel}(low)<0.03$. 

\item We do not see evidence of a shadow (on scales less than $\sim40$ kpc) of the UV background (from the H{\sc i} in the target galaxies) in a stack of our target galaxies.  This suggests that either the H{\sc i} columns are still large at 40 kpc scales or the majority of the far-UV background originates in the foreground of our sources (eg. $z<1$).\\

Given the dearth of ionizing emission from $z\sim1$ galaxies, future studies should focus on studying properties of $z\sim3$ galaxies in detail to determine what {\it why} some have large escape fractions.  In addition, it would be useful to identify unique characteristics of LyC-emitting galaxies at $z\sim3$ that can be used to identify LyC-emitters at even higher redshifts.   This would allow for a more accurate estimate of the contribution of star-forming galaxies to the high redshift ionizing background and help us understand the reionization process in more detail.


\end{itemize}

\acknowledgments

{\it Facilities:} \facility{Hubble}

\bibliographystyle{apj}
\bibliography{apj-jour,all_ref}

\end{document}